\documentclass[aps,prl,final,10pt,twocolumn,showpacs]{revtex4-1}

\usepackage{amsmath}
\usepackage{amsfonts}
\usepackage{amssymb}
\usepackage{graphicx}
\usepackage{subfigure}
\usepackage[pdfborder={0 0 0}]{hyperref}  
\usepackage{color}
\usepackage[latin1]{inputenc}
\usepackage{float}

\begin{document}

\title{Addendum to ``Anomalous scaling and super-roughness in the growth of CdTe polycrystalline films''}
\author{Fábio S. Nascimento}
\author{Angélica S. Mata}
\author{Silvio C. Ferreira}\email{silviojr@ufv.br}\thanks{On leave at Departament de Física i Enginyeria Nuclear, Universitat Politècnica de Catalunya, Spain.}
\author{Sukarno O. Ferreira}
\affiliation{Departamento de F\'{\i}sica, Universidade Federal de Vi\c{c}osa, 36571-000, Vi\c{c}osa - MG, Brazil}
\begin{abstract}
 The scaling of the growth of CdTe films on glass substrates was investigated by Mata \textit{et al.} [Phys. Rev. B \textbf{78}, 115305 (2008)]. Part of the analysis consisted of the estimation of the correlation length $\xi$ using the decay in the height-height correlation function. Afterwards, the dynamical exponent $z$ was determined using the scaling hypothesis $\xi\sim t^{1/z}$. In this Addendum, we show that the correlation lengths obtained by Mata \textit{et al.} provide a long wavelength coarsening exponent that does not correspond to the dynamical exponent $z$. We also show that the short wavelength coarsening exponent is consistent with the exponent $z$ obtained by Nascimento \textit{et al}. [arXiv:1101.1493] via generic dynamical scaling theory.
\end{abstract}

\maketitle

\textit{Introduction}. The scaling exponents involved in the deposition of CdTe on glass substrates has been investigated as functions of temperature \cite{Mata2008,Ferreira2006}. Details of the growth systems can be found in Ref. \cite{Ferreira2006}. The surfaces were scanned with a profiler and 1+1 interfaces of 300 microns and 4570 pixels of resolution were obtained. The scaling analysis in the real space (surface height fluctuations) involves the determination of four exponents: $\alpha$, $\beta$, $z$ and $\alpha_{loc}$, being only three of them independent due to the scaling relation $\alpha=\beta z$ \cite{Lopez}.  The exponents $\beta$ and $\alpha_{loc}$ were reported formerly in ref. \cite{Ferreira2006}.
Growth exponent $\beta$ was determined using the global interface width $W(t)$ defined as the rms fluctuations of profile height. The scaling law $W\sim t^{\beta}$ defines the growth exponent. The Hurst or local roughness exponent $\alpha_{loc}$, was determined using the rms height widths on a scale of length $\varepsilon$ and applying the scaling law  $w(\varepsilon)\sim \varepsilon^{\alpha_{loc}}$. 

In Ref. \cite{Mata2008}, it was investigated the dynamical exponent $z$  using the scaling law $\xi\sim t^{1/z}$. The correlation length $\xi$ was estimated using the decay in the two-point correlation function $\Gamma(\varepsilon,t)$. The roughness exponent $\alpha$ was thus obtained via scaling relation $\alpha=\beta z$. Two equivalent definitions of $\Gamma$ were used [Eqs. (7) and (8) in Ref. \cite{Mata2008}] and the exponents diverge less than 10\% \cite{Mata2008}.  The correlation length was finally obtained by solving $\int_0^\xi \Gamma d\varepsilon = 0.1\int_0^\infty\Gamma d\varepsilon$. Let us discuss the case in which $\Gamma$ is defined as the probability of the height difference between two points distant by $\varepsilon$ be lower than a fixed value $m=0.1|\tilde{h}_{max} - \tilde{h}_{min}|$, where $\tilde{h}_{max}$ and $\tilde{h}_{min}$ are the maximum and the minimum heights in the profile. The usual correlation $\Gamma=\langle h(x+\varepsilon)h(x)\rangle$ undergoes exactly the same effects that we shall discuss in this Addendum.

\begin{figure}[H]
 \centering
 \includegraphics[width=8.3cm]{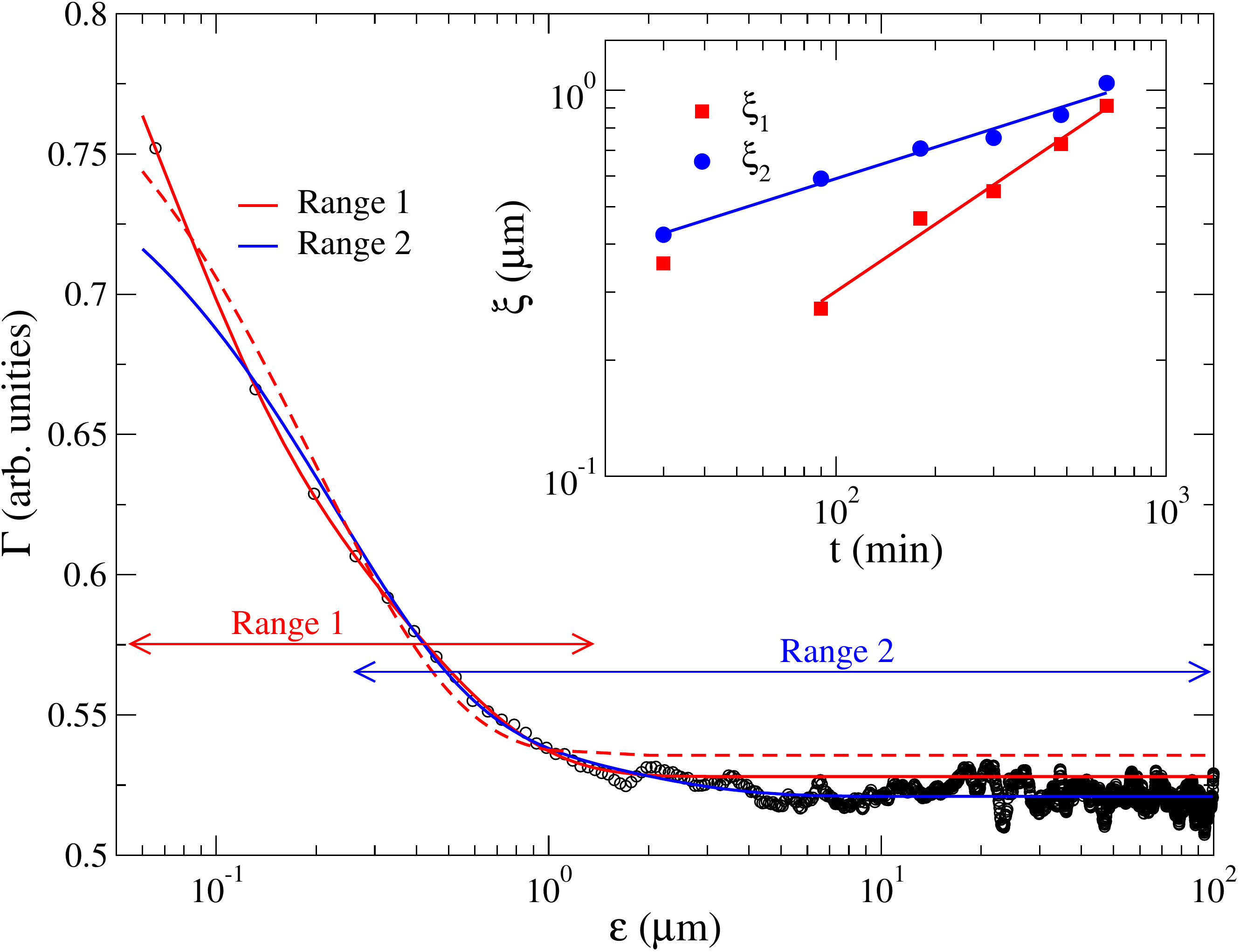}~
 \caption{(Color online) Height-height correlation function of the CdTe surfaces. The growth time and temperature were $90$ min and $300^\circ$C. Circles represent the experimental data while solid lines the two-exponential regressions in the indicated intervals. The dashed line is a single exponential regression in the range 1. Inset shows the correlation length obtained for the different regression intervals.}
 \label{fig:correl}
\end{figure}

\textit{Results}. A single exponential decay, commonly used to fit correlation functions, does not work properly for our data  as one can see in Fig. \ref{fig:correl}. Hence, a two-exponential decay was adopted [Eq. (9) in Ref. \cite{Mata2008}]. However, either a single or a double exponential provided the same exponents considering the error bars inasmuch as the respective correlation lengths are simply proportional. The region where the fit is done is indeed what matters. In Ref. \cite{Mata2008},  the tails corresponding to the long wavelength height fluctuations were included in the non-linear regressions and, when necessary, a few points at the shortest scales  were excluded (3 to 5 among more than $10^3$ points). Again, the exponents are not altered inside the margins of errors if these few points are or not included. However, if the tail is left out of the regression, the exponents may change considerably which is exactly the subject of this Addendum.

 In Fig. \ref{fig:correl}, we compare the regressions of the correlation functions in two ranges: discarding (range 1) and including (range 2) the tail.  The regression in the range 2, the same used in Ref. \cite{Mata2008}, misfits the data only for $\varepsilon \lesssim 0.2~\mu$m whereas the regression in the range 1 fits very well the small scales but deviates for $\varepsilon\gtrsim 2~\mu$m. The insertion to Fig. \ref{fig:correl} shows the correlation lengths against time and the respective power law regressions. Both cases yield quite satisfactory scaling laws with different exponents. For the regression in the regions 1 and 2 we obtained $\xi_1\sim t^{0.58}$ and $\xi_{2}\sim t^{0.27}$, respectively. Assuming $\xi_i\sim t^{1/z_i}$, where $i=1,2$, we found $z_1=1.72$ and $z_2=3.69$. Notice that the regressions provided two characteristic lengths of the same magnitude ($1<\xi_2/\xi_1\lesssim 2$) but exhibiting very different scalings with time. Repeating this analysis for lower temperatures ($T=150-250~^\circ$C) we found $\xi_1\lesssim 0.5~\mu$m  but no clear scaling law could be identified. We supposed that these scales are too small to be accurately analyzed using  a profiler with a large tip of 2.5 $\mu$m of radius. AFM images, in which the grain characteristic sizes do not exceed a few hundred nanometers,  confirm our hypothesis.  Only for $300~^\circ$C the grains are of the order of 1 $\mu$m as shown in Fig. 4b of Ref. \cite{Nascimento}.
 
 In Ref. \cite{Nascimento}, this system was investigated using the Generic Dynamical Scaling Theory (GDST), in which both real (height fluctuations) and momentum (power spectra) spaces analysis are performed \cite{Lopez}. It was confirmed that the scaling exponents obtained in \cite{Mata2008}, which corresponds to range 2 in Fig. \ref{fig:correl}, do not agree with those of the GDST. However, the dynamical exponent obtained via GDST, $z=1.77\pm 0.05$, is consistent with the estimate $z_1=1.72$ for $T=300~^\circ$C, indicating that the dynamical exponent $z$ corresponds to the coarsening exponent $z_1$. 
 
 \begin{figure}[H]
 \centering
 \includegraphics[width=8.3cm]{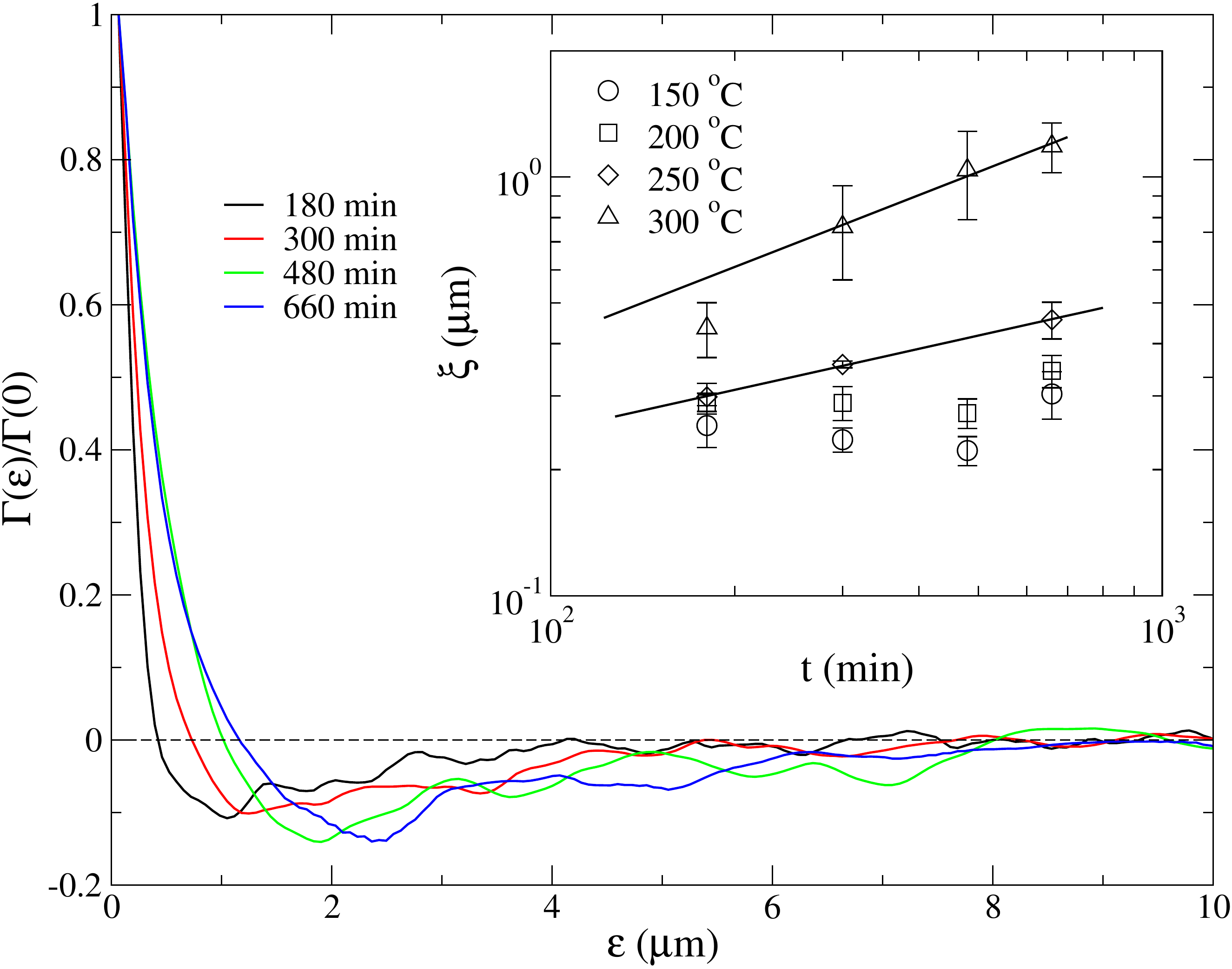}~
 \caption{(Color online) Slope-slope correlation functions for the CdTe surfaces. In the main plot, the correlation for distinct times and $T=300~^\circ$C are shown. Inset shows the correlation lengthy against time. Straight lines are power law regressions.}
 \label{fig:corgrad}
\end{figure}

 Finally, we measured the slope-slope correlation function $\Gamma(\varepsilon)= \langle \nabla h(x+\varepsilon) \nabla h(x)\rangle$ that was not investigated in Ref. \cite{Mata2008}. The correlation length can be defined as the first zero of the correlation function ($\Gamma(\xi)=0$) and, consequently, no regression is required to estimate $\xi$. The correlation functions and lengths are shown in Fig. \ref{fig:corgrad}. Similarly to the height-height correlation, there is no satisfactory power law regime for $T=150~^\circ$C and $200~^\circ$C. For $250~^\circ$ and $300~^\circ$C, the slopes provides $1/z = 0.32$ and $1/z = 0.57$, respectively, which are consistent with GDST analysis in \cite{Nascimento}.   
   
\textit{Conclusions}. The conclusions of Ref. \cite{Mata2008} are partially modified since we cannot apply GDST to infer about which dynamical scaling regime the system belongs to. The regime claimed as intrinsically anomalous, in which $\alpha<1$ and $\alpha\ne\alpha_{loc}$, has  been shown to be described by the exponents $\alpha_{loc}=1$ and $\alpha>1$ \cite{Nascimento}. The exponent $z$ presented in \cite{Mata2008} is not the dynamical exponent of the GDST. It would be better to refer to this exponent as a long wavelength coarsening exponent, which increases with temperature whereas the actual dynamical exponent obtained via GDST in Ref. \cite{Nascimento} decreases. 

These results also stress out the difficulties usually observed in the fitting procedure of experimental data and the problems which can arise from a misinterpretation of the parameters obtained. For instance, the power spectrum itself used in GDST is usually noisy due the Fourier transforms. The GDST proposes that the power spectra at different growth times collapse in a single curve if the suitable rescaling is applied \cite{Lopez}. Even though high quality collapses of the power spectra in experimental data have been reported \cite{Nascimento,Cordoba}, there are also examples where the large noise does not allow a clear resolution of the collapses \cite{Bru,Soriano}. Therefore, a combination of different methods of scaling is the best procedure to avoid misinterpretations in the scaling analyses.

\begin{acknowledgments}
 This work was supported by the Brazilian agencies CNPq, FAPEMIG and CAPES. SCF thanks the kind hospitality at the Departament de Física i Enginyeria Nuclear/UPC.
\end{acknowledgments}

\end{document}